\documentclass[aps,prl,twocolumn,superscriptaddress,preprintnumbers]{revtex4}

\usepackage{epsf,epsfig,amsmath,amssymb,amsfonts}
\usepackage{color}
\usepackage{slashed}
\usepackage{stmaryrd}

\newsavebox{\ns}
\newsavebox{\dbrane}

\def\be{\begin{equation}}
\def\ee{\end{equation}}
\def\bea{\begin{eqnarray}}
\def\eea{\end{eqnarray}}

\def\Dslash{\,\,{\raise.15ex\hbox{/}\mkern-12mu D}}
\def\Dbarslash{\,\,{\raise.15ex\hbox{/}\mkern-12mu {\bar D}}}
\def\delslash{\,\,{\raise.15ex\hbox{/}\mkern-9mu \partial}}
\def\delbarslash{\,\,{\raise.15ex\hbox{/}\mkern-9mu {\bar\partial}}}
\def\pslash{\,\,{\raise.15ex\hbox{/}\mkern-9mu p}}
\def\calDslash{\,\,{\raise.15ex\hbox{/}\mkern-12mu {\cal D}}}

\newcommand\diff{\mbox{d}}

\newcommand{\vol}{\mbox{vol}}

\newcommand{\nn}{\nonumber \\}

\newcommand{\dd}{\diff}

\def\a{\alpha}

\def\b{\beta}

\allowdisplaybreaks

\begin{document}

\title{$I$ in generalized supergravity}

\preprint{APCTP Pre2017 - 015}          

\preprint{KUNS-2696}

\preprint{IPM/P-2017/028}
\vskip 1 cm

 \author{T. Araujo}
 \affiliation{Asia Pacific Center for Theoretical Physics, Postech, Pohang 37673, Korea}
 \author{E. \'O Colg\'ain}
 \affiliation{Asia Pacific Center for Theoretical Physics, Postech, Pohang 37673, Korea}
\author{J. Sakamoto}
 \affiliation{Department of Physics, Kyoto University,  Kitashirakawa, Kyoto 606-8502, Japan}
\author{M. M. Sheikh-Jabbari}
\affiliation{School of Physics, Institute for Research in Fundamental Sciences (IPM), P.O.Box 19395-5531, Tehran, Iran}
\author{K. Yoshida}
 \affiliation{Department of Physics, Kyoto University,  Kitashirakawa, Kyoto 606-8502, Japan}

\begin{abstract}
\noindent 
We showed in previous work that for homogeneous Yang-Baxter (YB) deformations of AdS$_5\times$S$^5$, the open string metric and coupling, and as a result the closed string density $e^{-2 \Phi} \sqrt{g}$, remain undeformed. In this work, in addition to extending these results to the deformation associated with the modified CYBE, or $\eta$-deformation, we identify the Page forms as the open string counterpart for RR fields and demonstrate case by case that the non-zero Page forms remain invariant under YB deformations. We give a physical meaning to the Killing vector $I$ of generalized supergravity and show for all YB deformations: 1) $I$ appears as a current for center of mass motion on the worldvolume of a D-branes probing the background, 2)  $I$ is equal to the divergence of the noncommutativity parameter,  3)  $I$ exhibits ``holographic'' behavior, where the radial component of $I$ vanishes at the AdS boundary, and 4)  in pure spinor formalism $I$ is related to a certain state in the BRST cohomology.

\end{abstract}

\maketitle

\setcounter{equation}{0}

\section{Introduction} \label{Introduction}

Over recent years we have witnessed an explosion in interest in integrability preserving deformations of AdS$_5\times$S$^5$ superstring, collectively dubbed Yang-Baxter (YB) deformations \cite{Klimcik:2002zj,Klimcik:2008eq, Delduc:2013fga, Matsumoto:2015jja}, which includes T-duality shift T-duality (TsT) transformations \cite{Lunin:2005jy} as a special case. As a by-product of this breakthrough, a large family of supergravity solutions, or geometries, of varying complexity can be defined. In \cite{Araujo:2017jkb, Araujo:2017jap}, it was observed that within the ensuing geometric madness, there was a simplifying principle: for homogeneous YB deformations it was shown that in open string frame the metric and string coupling were undeformed and all information about the deformation resided in a noncommutativity (NC) parameter, or antisymmetric bivector $\Theta$, thus unlocking a key to the AdS/CFT interpretation (see also \cite{vanTongeren:2015uha, vanTongeren:2016eeb}). In this letter, noting the key role played by pure gauge contributions to the NSNS two-form, we extend the results to $\eta$-deformations.

Concurrently, significant progress has been made in the Green-Schwarz (GS) formulation of superstring theory. From old work \cite{Grisaru:1985fv}, the on-shell condition of type II supergravities ensures the invariance of the GS string action under kappa-symmetry (which is a local fermionic symmetry and necessary to get the right number of physical spacetime fermions). On the other hand, the reverse statement had not been clarified even after more than three decades. In groundbreaking work by Tseytlin and Wulff \cite{Wulff:2016tju}, it has been shown that the kappa-symmetry of the GS string action defined on an arbitrary background leads to {\it generalized} type II supergravities \cite{Arutyunov:2015mqj} including an extra Killing vector field $I$, rather than the usual, well-known supergravities. 
The derivation of generalized supergravity is robust and the result is so striking that the fundamental aspects of superstring theory must be carefully reconsidered. Hence, the study of generalized type II supergravities is a fundamental issue concerned with the formulation of superstring theory. Here, we confine our attention to generalized type IIB supergravity. We reproduce the equations of motion (EOM) of the NSNS sector in the appendix. 

It is worth making a comment on the Weyl invariance of  string theory when the target space is given by a solution of the generalized supergravity. In the original paper \cite{Arutyunov:2015mqj}, 
the world-sheet theory was described as non-Weyl invariant. Then in the recent work \cite{Sakamoto:2017wor}, 
the modified Fradkin-Tseytlin (FT) term was proposed and the Weyl invariance was shown in the doubled formalism, where the counter-term is non-local due to the dual coordinate dependence of the dilaton. 
But we recall here that the interpretation of the FT term itself is different from that of the usual counter-term. In particular, it breaks the classical Weyl invariance. However, this issue can be overcome 
because the FT term should be regarded as a one-loop 
quantum correction rather than the classical one. The non-locality of the modified FT term can also be avoided with the same logic. In fact, the extra vector field $I$ is identified with a trace of non-geometric $Q$-flux  as as shown in \cite{Sakamoto:2017cpu}, 
and hence the solutions of the generalized supergravities are non-geometric. To account for the global property of the non-geometric background, the non-locality of the FT term should be rather natural. 

In generalized type IIB supergravity itself, there are a lot of open questions, including, \textit{what is the physical interpretation of $I$?} In order to tackle these issues, it is useful to employ the YB deformation because this can be regarded as a solution generation technique in generalized supergravity. Although we appreciate the profound derivation of the generalized type IIB supergravity from the kappa-invariance nowadays, it is remarkable that its discovery was a by-product of an enriched understanding of integrable deformations of the AdS$_5\times$S$^5$ superstring. 

The integrable deformation technique that concerns us was invented by Klimcik for the principal chiral models based on the modified classical Yang-Baxter equation (mCYBE) \cite{Klimcik:2002zj,Klimcik:2008eq}. According to this procedure, an integrable deformation is specified by taking a classical $r$-matrix satisfying the mCYBE and hence this procedure is often called the Yang-Baxter deformation. Klimcik's work was subsequently generalized to the symmetric coset case \cite{Delduc:2013fga} and the homogeneous CYBE \cite{Matsumoto:2015jja}. Since the classical action of the AdS$_5\times$S$^5$ superstring \cite{Metsaev:1998it} has a construction based on the supercoset 
\begin{eqnarray}
\frac{PSU(2,2|4)}{SO(1,4) \times SO(5)}, 
\end{eqnarray}
the $\mathbf{Z}_4$-grading of the superalgebra leads to classical integrability \cite{Bena:2003wd}. 
Thus, the YB deformation is also applicable to the AdS$_5\times$S$^5$ superstring theory \cite{Delduc:2013qra,Kawaguchi:2014qwa}. The pioneering work of Delduc, Magro and Vicedo \cite{Delduc:2013qra}, based on the mCYBE, is the standard $q$-deformation of $\mathfrak{su}(2,2|4)$ with the classical $r$-matrix of Drinfeld-Jimbo type. 
This is often called the $\eta$-deformation. After performing supercoset construction, the full background, including the RR sector, was worked out \cite{Arutyunov:2015qva}, which led to the surprise that the resulting background was not a solution of the usual type IIB supergravity, but a generalized variant \cite{Arutyunov:2015mqj}. 
 
To date, a considerable amount of community effort has focused on the largely technical exercise of mapping out the dictionary between $r$-matrices and geometries, or YB deformations, including connections to (non-Abelian) T-duality \cite{Matsumoto:2014nra, Matsumoto:2014gwa, Matsumoto:2014ubv, Matsumoto:2015uja, Kawaguchi:2014fca, vanTongeren:2015soa, Osten:2016dvf, Hoare:2016wca, Kyono2016jqy, Hoare:2016ibq, Hoare:2016hwh, Orlando:2016qqu, Borsato:2016ose}. Following the latter thread further, the relation between Double Field Theory (DFT) \cite{Siegel:1993th, Siegel:1993xq, Hull:2009mi} was elucidated in \cite{Sakatani:2016fvh,Baguet:2016prz,Sakamoto:2017wor, Sakamoto:2017cpu} and  some implications for the AdS/CFT dual gauge theory, including whether planar integrability is preserved \cite{Beisert:2005if, Guica:2017mtd}, have been teased out in \cite{Araujo:2017jkb, Araujo:2017jap, vanTongeren:2015uha, vanTongeren:2016eeb, Roychowdhury:2017oed, Roychowdhury:2017vdo}. For studies of integrability, including classical solutions on the $\eta$-deformed background, see \cite{AM-NR,Kame-coord,magnon,Bozhilov:2016owo, Arutyunov:2016ysi, Ahn:2016egk, Hernandez:2017raj, Klabbers:2017vtw}. In this letter, we return to the physical interpretation of the Killing vector $I$. 

We recall for homogeneous YB deformations that the Killing vector $I$ is related to the divergence of the noncommutativity (NC) parameter, or bivector $\Theta$ \cite{Araujo:2017jkb, Araujo:2017jap}, 
\be
\label{div_theta}
I^{N} = \nabla_{M} \Theta^{MN} ,  
\ee 
where $\nabla$ is the covariant derivative with respect to the open string metric. The validity of this relation has been checked for homogeneous YB deformations on a case by case basis and an outline of its derivation was given in \cite{Araujo:2017jap}. In this letter, we extend this result to the known $\eta$-deformations \cite{Delduc:2013qra, Arutyunov:2013ega, Delduc:2014kha,Arutyunov:2015qva} and provide a derivation in terms of D3-branes probing the background, where $I$ may be interpreted as a current for the center of mass motion on the D3-brane worldvolume. 

The conclusion that (\ref{div_theta}) holds for all YB deformations, including $\eta$-deformations, hinges on the observation that Page forms \cite{Page}, special combinations of RR fluxes and NSNS two-form that appear in the D-brane action, facilitate a simple rewriting of the equations of generalized supergravity and that the Page five-form associated to the D3-brane is invariant, which we have confirmed on a case by case basis. Recalling that the corresponding quantized Page charges are mapped to the ranks of gauge groups in the dual field theory, one arrives at some intuitive understanding for this invariance through AdS/CFT. More generally, we conjecture that the non-zero Page forms are invariants of YB deformations. In addition, through complementary comments on the role of $I$ as a certain state in the BRST cohomology in pure spinor formalism, our work provides $I$ with a more physical flavor. 

In the remainder of the letter, we study $\eta$-deformations. Concretely, we demonstrate that the appropriate inclusion of pure gauge contributions to the NSNS two-form indeed results in an open string metric that is undeformed. For the explicit Arutyunov, Borsato, Frolov (ABF) solution \cite{Arutyunov:2013ega, Arutyunov:2015qva}, we check that the well-known T-duality invariant $e^{-2 \Phi} \sqrt{g}$ is also an invariant of the deformation. This extends an earlier observation, originally only valid for homogeneous YB deformations \cite{Araujo:2017jap}, to YB deformations associated to $\eta$-deformations.  We use this fact and (\ref{div_theta}) to determine the NSNS sector of $\eta$-deformations of AdS$_5 \times$S$^5$ \cite{Delduc:2014kha}\footnote{In \cite{Delduc:2014kha} only the metric and NSNS two-form are computed. 
The remainder of the solution, including the dilaton and RR sector can be determined by performing the supercoset construction directly, as explained in \cite{Arutyunov:2015qva,Kyono2016jqy}. Alternatively, it can also be fixed by analytic continuations \cite{Hoare:2016ibq}, starting from the original $\eta$-deformed background \cite{Arutyunov:2013ega,Arutyunov:2015qva}.}, in the process finding that the open string metric is undeformed, in line with our expectations. We check all expressions against the dilaton EOM, which we show is satisfied in all cases.

\section{Open string frame}

In the context of string theory and its related effective field theories, depending on the probes used and the associated degrees of freedom, one may conveniently describe the system in different frames. These frames are related by field redefinitions and carry the name of the associated probe. For example, closed strings probe string frame and particles probe Einstein frame. 
Within string theory, especially when we have D-brane probes, or when we have spacetimes with a causal boundary like AdS geometry, one may choose to probe spacetime with D-branes or ``open strings''.  It is known, especially after the seminal work of Seiberg and Witten \cite{Seiberg:1999vs}, that the open string  and closed string frames in the NSNS sector are different if we have a Kalb-Ramond $B$-field in the background. In this section, we first review the open string frame in the NSNS sector and then introduce the open string frame fields for the RR sector.

\subsection{Open string frame, NSNS sector}

Given a triple ($g_{MN}, B_{MN}, \Phi$) comprising the closed string metric $g$, closed string coupling $g_s e^{\Phi}$ and NSNS two-form $B$, one can define an open string metric $G$, an antisymmetric bivector $\Theta$ and open string coupling $G_s$ \cite{Seiberg:1999vs} \footnote{See \cite{Duff:1989tf} for an earlier incidence of open string metric and coupling.}: 
\bea
G_{MN} &=& \left( g- Bg^{-1}B\right)_{MN},\label{Open-metric} \\
\Theta^{MN} &=& - \left( ({g + B})^{-1} B ({g - B})^{-1} \right)^{MN},\label{Theta-def} \\
 G_s &=& g_s {\rm e}^{\Phi} \left( \frac{\det (g+B)}{ \det g} \right)^{\frac{1}{2}} \label{Gs}.
\eea
The above are called open string fields because they are the combinations naturally appearing in the DBI part of the brane action describing open strings. One can also directly show that these are the fields appearing in the low-energy effective limit of open string scatterings \cite{Seiberg:1999vs, Open-string-scattering}. 

For $r$-matrix solutions to the homogeneous CYBE it was noted in \cite{Araujo:2017jkb, Araujo:2017jap} that following the frame change, the open string metric is simply undeformed AdS$_5$ with constant open string coupling $G_s = g_s$ and all information about the YB deformation resides in the bivector $\Theta$. Invariance of the open string metric and coupling under $O(d, d)$ transformations had been observed earlier \cite{Berman:2000jw}. In addition, the bivector exhibits holographic commutativity, namely $\Theta$ depends on the holographic direction $z$, but all dependence drops out at the boundary $z=0$. To be more precise, it was noted that $\Theta^{\mu z} |_{z=0} = 0$. At $z=0$ one is expected to make contact with a NC deformation of $\mathcal{N} =4$ super Yang-Mills: in support of this correspondence, it has been shown case by case that the bivector, evaluated at $z=0$, agrees with the Moyal bracket arising from the Drinfeld twist \cite{Drinfeld} of the conformal algebra. At the level of algebra, this establishes an equivalence between homogeneous YB deformations and conformal twists, thereby substantiating an earlier conjecture of van Tongeren \cite{vanTongeren:2015uha}. 

A corollary of these results is the relation \cite{Araujo:2017jap}, 
\be\label{G-Phi-g}
\sqrt{\det G} = e^{-2 \Phi} \sqrt{\det g}. 
\ee 
Since $G$ is undeformed AdS$_5 \times$S$^5$ metric, this implies that the density $e^{-2 \Phi} \sqrt{\det g}$, in addition to being a well-known invariant of Buscher's T-duality, is more generally an invariant of YB deformations based on $r$-matrix solutions to the homogeneous CYBE. We will show later that this extends to $\eta$-deformations \cite{Delduc:2013qra, Arutyunov:2013ega, Delduc:2014kha}.  

It was shown in \cite{Araujo:2017jkb, Araujo:2017jap} that for homogeneous YB deformations we always have equation \eqref{div_theta}, thus relating $I$ to the open string bivector $\Theta$. This relation was explicitly checked for an exhaustive set of backgrounds associated with conformal twists and moreover, we gave a symmetry based argument for \eqref{div_theta}, but with an incomplete derivation. In this letter, we complete the derivation and demonstrate below that \eqref{div_theta}, as well as the symmetry based argument leading to it, remains valid for the mCYBE case associated with $\eta$-deformations. In addition, we show that an expression for the dilaton can be read off from (\ref{G-Phi-g}), once one uses the fact that $G$ is undeformed. 

\subsection{Open string frame, RR sector}

In closed string theory we have RR form fields too and one may wonder what are their corresponding fields in the open string frame. Similarly to the NSNS sector, one may look again into the D-brane action, but now the Chern-Simons or Wess-Zumino part of it which contains RR-forms. Here we will conveniently denote this part of a $p$-brane action by $S_{WZ}$, which, when the field strength of the gauge field on the brane is set to zero, takes the form:
\be\label{WZ}
S_{WZ}=\int_{\Sigma_p} \sum_{n\leq p+1} C_n\wedge e^B,
\ee
where $C_n$ are $n$-form RR fields pulled back to the Dp-brane worldvolume $\Sigma_p$. 

The above suggests that combinations of RR-forms and $B$-field are appropriate for open string frame RR forms.  When $B$ is a constant exterior derivative (``field strength''), the above sum is related to a $p+2$-form Page form  \cite{Page}: 
\bea
Q_1 &=& F_1, \quad Q_3 = F_3 + B F_1, \quad Q_5 = F_5 + B F_3 + \frac{1}{2} B^2 F_1, \nn
Q_7 &=&  - * F_3 + B F_5 + \frac{1}{2} B^2 F_3 + \frac{1}{3!} B^3 F_1, \label{Page-forms}\\
Q_9 &=& * F_1 - B * F_3 + \frac{1}{2} B^2 F_5 + \frac{1}{3!} B^3 F_3 + \frac{1}{4!} B^4 F_1, \nonumber
\eea
where we have omitted wedge products and employed $B^2 = B \wedge B$, etc. Closure of $Q_n$ is guaranteed by the Bianchis/EOMs of usual type IIB supergravity. We take Page forms \eqref{Page-forms} as the open string frame fields corresponding to RR forms.

One may view the pull back of the above Page forms on the brane  as ``Page charge densities'' and define  Page charges \cite{Page} as  integrals of them over compact cycles (where the branes wrap).  Page charges have been shown to be 
localized, conserved and most importantly quantized \cite{Marolf:2000cb}, so that they may be identified with the rank of gauge groups in the dual theory.

\section{Page forms and generalized supergravity} 
It is known that homogeneous YB deformations, which are equivalent to (generalized) T-dualities \cite{Hoare:2016wsk, Borsato:2016pas,Borsato:2017qsx}, result in supergravity solutions provided a unimodular condition holds \cite{Borsato:2016ose}. For non-unimodular YB deformations one finds a solution to generalized supergravity \cite{Arutyunov:2015mqj}, where, as explained in the appendix, the EOMs involve an extra vector field $I$, which is a Killing vector of the original and the deformed background. The full set of field equations of generalized supergravity may be found in \cite{Wulff:2016tju,Arutyunov:2015mqj}. 

Here, we make the observation that the Bianchis and EOMs of generalized supergravity simplify considerably once expressed in terms of Page forms \eqref{Page-forms}.
When $I \neq 0$, the equations of generalized supergravity read \cite{Arutyunov:2015mqj},
\bea
\label{F1_mod} \dd F_1 &=& i_{I} F_3 + i_{I} B \wedge F_1, \\ 
\label{F3_mod} \dd F_3 + H \wedge F_1 &=& i_{I} F_5 + i_{I} B \wedge F_3, \\
\label{F5_mod} \dd F_5 + H \wedge F_3 &=& - i_{I} * F_3 + i_{I} B \wedge F_5, \\
\label{F7_mod} \dd * F_3 - H \wedge F_5 &=& - i_{I} * F_1 + i_{I} B \wedge * F_3, \\
\label{F9_mod} \dd * F_1 - H \wedge * F_3 &=& i_{I} B  \wedge * F_1.  
\eea

Recast in terms of the Page forms (\ref{Page-forms}), the modified equations take a remarkably simple form 
\bea
\label{Q_mod} \dd Q_1 &=& i_{I} Q_3, \quad 
 \dd Q_3 = i_{I} Q_5, \quad \dd Q_5 =  i_{I} Q_7,  \nn 
\quad \dd Q_7 &=&  i _{I} Q_9. 
\eea
We omit $\dd Q_9$ due to its length. It is now clear that the Page forms are no longer closed. Nevertheless, the above are a consistent set of equations if $I$ is a Killing vector field. 

To get a better feel for the rewriting, let us explicitly demonstrate how the first two equations in (\ref{Q_mod}) come about. Similar logic applies to the later equations. From $i_{I} F_1 = 0$, which is a property of all generalized supergravity solutions \cite{Arutyunov:2015mqj}, we see that (\ref{F1_mod}) is simply the first equation in (\ref{Q_mod}). Furthermore, combining (\ref{F1_mod}) and (\ref{F3_mod}), we have: 
\bea
\dd ( F_3 + B F_1) &=& i_{I} F_5 + i_{I} B  F_3 + B (  i_{I} F_3 + i_{I} B  F_1), \nn
&=& i_{I} F_5 + i_{I} (B  F_3) + \frac{1}{2} i_{I} (B^2 F_1)  \nn
&=& i_{I} Q_5, 
\eea
which explains the second equation in (\ref{Q_mod}). 

It is easy to convince oneself through explicit calculation, either case by case or employing a general ansatz, that the \textit{non-zero} Page forms and associated Page charges are invariants of TsT transformations. For more general YB deformations, this can be checked for explicit solutions case by case. In particular, we have checked all examples presented in \cite{Orlando:2016qqu}, where $I \neq 0$. This leads us to the conjecture:
\begin{quote}
{\it Non-zero Page forms and associated Page charges  are invariants of all Yang-Baxter deformations.}
\end{quote}
The above extends our earlier result for invariance of open string metric and coupling to the RR-sector.

At an intuitive level, this statement is somewhat obvious and expected: in the current setting, we start with AdS$_5\times$S$^5$, a geometry sourced exclusively by D3-branes, we perform a special transformation that introduces a continuous constant deformation parameter, which is  \textit{a priori} not quantized. In this light, it is reasonable to expect that the non-zero Page forms and charges are invariant. 

It is instructive to illustrate our point in the simplest example of the Hashimoto-Itzhaki, Maldacena-Russo geometry \cite{Hashimoto:1999ut, Maldacena:1999mh} (see also \cite{Alishahiha:1999ci}), 
\bea\label{MR-solution}
\dd s^2 &=& \frac{(-\dd t^2 + \dd x_3^2 + \dd z^2)}{z^2} + \frac{z^2( \dd x_1^2 + \dd x_2^2)}{(z^4 + \eta^2)}  + \dd s^2(S^5), \nn
B &=& \frac{\eta}{z^4 + \eta^2} \dd x_1 \wedge \dd x_2, \quad e^{ 2 \Phi} = \frac{z^4}{(z^4 + \eta^2)}, \nn
F_3 &=&  \frac{4 \eta }{z^5} \dd t \wedge \dd x_3 \wedge \dd z,  \\
F_5 &=&  ( 1+ *)\frac{4}{z (z^4 + \eta^2)} \dd t \wedge \dd x_1 \wedge \dd x_2 \wedge \dd x_3 \wedge \dd z. \nonumber
\eea
A short calculation reveals that the only non-zero Page forms associated with this geometry are 
\be\label{Q3-Q5-MR-example}
Q_3 = \frac{4 \eta }{z^5} \dd t \wedge \dd x_3 \wedge \dd z, \quad Q_5 = (1+*) 4 \vol(AdS_5).
\ee
The important point to take away from this example is that $Q_5$ has not changed in the deformation and is simply the original undeformed five-form flux. While we defer the general proof to an upcoming publication, one may check by direct computation for an exhaustive list of examples  that $Q_5$ is the undeformed five-form flux for all YB deformations, homogeneous or modified, of AdS$_5\times$S$^5$ (for $\eta$-deformations associated with mCYBE this can be checked using the ABF solution \cite{Arutyunov:2013ega}). Our conjecture applies equally to YB deformations of different geometries, where different Page forms will be invariant. 

\section{Physical meaning of $I$}
In this section, we arrive at the main point of this letter, namely to address the physical meaning of the Killing vector $I$ of generalized supergravity. We will now give two complementary perspectives: we will identify $I$ as a current on the D-brane worldvolume and elucidate its meaning in the pure spinor formalism of the string $\sigma$-model. Our treatment improves on earlier analysis \cite{Araujo:2017jap} by specifying the D-brane, in this case a D3-brane, and illustrating how it couples to the background through a Wess-Zumino term.

\subsection{$\Lambda$-symmetry and $I$ on the brane}\label{sec:Lambda-sym}

In \cite{Araujo:2017jkb, Araujo:2017jap} we argued that \eqref{div_theta} may be derived from a symmetry principle, the $\Lambda$-symmetry: (generalized) supergravity equations are invariant under the $\Lambda$-symmetry,
\be\label{Lambda-closed}
B \rightarrow B + \dd \Lambda, 
\ee
where $\Lambda$ is an arbitrary one-form. At the level of generalized gravity this is manifest because the EOMs depend only on the  field strength $H = \dd B$ (see appendix B of \cite{Araujo:2017jap}). In the presence of open strings or D-branes, however, we have a $U(1)$ gauge field on the brane, $A$, with field strength $F=\dd A$.
If we have a stack of $N$ coincident D-branes this $U(1)$ symmetry gets enhanced to $U(N)$ \cite{Witten:1995im} and one may decompose it into $SU(N)$ fields and the ``center of mass'' $U(1)$ part. In what follows we set the $SU(N)$ part of the gauge fields to zero, since it does not mix with the closed string fields, and only focus on this center of mass $U(1)$ part. 

In the presence of D-branes the $\Lambda$-symmetry \eqref{Lambda-closed} is replaced with \cite{Witten:1995im, SheikhJabbari:1999ba}
\be\label{Lambda-open}
B \rightarrow B + \dd \Lambda, \qquad A\rightarrow A-\Lambda.
\ee
Then, besides $H$ there is another $\Lambda$-gauge invariant combination
\be
{\cal F}=B+F=B+\dd A,
\ee
and one can check that the DBI part of the brane action only involves ${\cal F}$ \cite{Leigh:1989jq}:
\be
S_{DBI} = \int_{\Sigma_p} \frac{e^{- \Phi}}{g_s} \sqrt{\det ( g + {\cal F})}. 
\ee
We comment in passing that ${\cal F}$ and hence the DBI action, is invariant under the $U(1)$ center of mass gauge symmetry, the $\lambda$-symmetry, under which $B$ remains invariant and $A\rightarrow A+\dd \lambda$, where $\lambda$ is an arbitrary scalar function.  

The NC description for the DBI action with the open string field $\Theta$ as NC parameter, as considered in \cite{Seiberg:1999vs}, is only a valid description in the specific $\Lambda$-gauge $\Lambda=A$. In this gauge ${\cal F}=B$.  Hereafter, we will be working in this $\Lambda$-gauge. In fact, the closed string-open string frame map \eqref{Open-metric} - \eqref{Gs}, expressed in this gauge \cite{Seiberg:1999vs}, yields 
\be\label{seiberg_witten}
\frac{1}{G_s} \sqrt{\det G} = \frac{e^{-\Phi}}{g_s} \sqrt{\det (g+B)}.
\ee

We will prove below that \eqref{div_theta} is a consequence of $\Lambda$-symmetry, of course once we fix the ${\cal F}=B$ gauge. To this end, we first note that $M,N$ indices are bulk AdS$_5$ indices, while the fields appearing in the brane action are pullbacks on the brane worldvolume. So, we may consider a brane that is AdS-filling, for example a D7-brane wrapping three dimensions on the 
S$^5$ piece. Alternatively, we may consider a D3-brane on the AdS$_5$ part, and recall that the brane worldvolume is codimension-one on this 5D spacetime. 

We shall choose the latter, where it is easy to show that if the D3-brane has no expansion along the radial direction in Poincar\'e patch of AdS$_5$, parameterized by $z$ coordinate, we have
\be\label{DBI-action}
S_{DBI} = 4\int_{\text{AdS}_5} \frac{e^{- \Phi}}{g_s} \sqrt{\det ( g + B)}=4\int_{\text{AdS}_5} \frac{1}{G_s} \sqrt{\det G},
\ee
where we used \eqref{seiberg_witten}. In addition, we used the facts that open string metric is simply the undeformed AdS$_5$ metric, $G_s$ is a constant, and the factor of 4 has appeared as a result of exterior differentiation of the D3-brane DBI, which produces a 5D action from a 4D action. Our treatment here mirrors similar analysis in  \cite{Schwarz:2013wra}.  

Once again we point out that the action \eqref{DBI-action} is written in a specific $\Lambda$-gauge. A consistency condition of fixing this gauge, recalling \eqref{Lambda-open}, implies
\be
\label{Lambda-A-variation}
\delta_\Lambda S_{DBI}=-\delta_AS_{DBI},
\ee
and hence the EOM for the gauge field $A$ is related to the $\Lambda$ invariance of the $\Lambda$-gauge fixed action. We shall return to this point momentarily. 

The full D3-brane action besides the DBI part, also involves a Wess-Zumino part,  which for a D3-brane in ${\cal F}=B$ $\Lambda$-gauge may be written as
\be
\label{WZ-D3}
S_{WZ} =   - \int_{\text{AdS}_5}  ( F_5 + B \wedge F_3 + \frac{1}{2} B^2 \wedge F_1).
\ee
As a side comment, we note that with the factor 4 in the DBI part \eqref{DBI-action},  we get the ``no force" condition cancellation of DBI and the WZ parts for flat, undeformed D3-branes. 

Now, we vary the overall action $S_{DBI}+S_{WZ}$ with respect to $\Lambda$. From the variation of the DBI term, we get \cite{Araujo:2017jap}
\be
\delta_{\Lambda} S_{DBI} =  - 4 \frac{\sqrt{\det G}}{G_s} \nabla_{M} \Theta^{MN} \delta \Lambda_N, 
\ee
where we have used (\ref{seiberg_witten}) and $\Theta^{MN} = (g + B)^{[MN]}$, whereas from the variation of $S_{WZ}$ we get 
\bea
\delta_{\Lambda} S_{WZ} &=&  -  ( F_3 + B \wedge F_1 ) \wedge \dd \delta \Lambda \nn &=&  - i_{I} Q_5 \wedge \delta \Lambda, \\
&=& -  \frac{4}{g_s}  i_{I} \vol(AdS_5) \wedge \delta \Lambda =  \frac{4}{g_s} \sqrt{\det G} I^{N} \delta \Lambda_{N}.  \nonumber
\eea
In the second line, we have used (\ref{Q_mod}) and in the third line, we have employed 
\be
\label{Q5}
Q_5 = \frac{4}{g_s} [ \vol(AdS_5) + \vol (S^5) ], 
\ee
where we have reinstated $g_s$. Recalling that $G_s = g_s$, combining the two contributions, we arrive at (\ref{div_theta}).  In other words, \eqref{div_theta} is nothing but the EOM for the gauge field $A$ in the $\Lambda=A$ gauge, or in other words, \eqref{div_theta} is the condition for consistency of $\Lambda=A$ gauge fixing. We stress that the $\lambda$-symmetry is still unfixed and that our arguments above do not depend on $\lambda$-symmetry gauge fixing. \footnote{We should note that expressions for the Page forms \eqref{Page-forms} are not $\Lambda$-gauge invariant, as they are written in a specific $\Lambda$-gauge. This $\Lambda$-gauge is the same as the one in which we have written our closed string to open string map. Moreover, this is the same gauge in which the $Z_M$ vector field of generalized supergravity \eqref{Z-solution} is written. We thank Ben Hoare for a clarifying discussion on this point.}

Given the above result, one may ask if it can be written in a different $\Lambda$-gauge. The answer is simply yes: if we fix the $B=0$ gauge, using \eqref{Lambda-A-variation}, one can readily see that the EOM for the gauge field $A$ of the action
\be
S=\int_{\Sigma_4} \frac{1}{G_s}\sqrt{G+F}- A_\mu I^\mu
\ee
yields the pullback of \eqref{div_theta} on the brane worldvolume.  The above action is interesting because it provides a new perspective on the vector field $I$: $I$ is a current coupled to the center of mass $U(1)$. 

Our derivation above  is true for all YB deformations of AdS$_5\times$S$^5$ as long as (\ref{Q5}) holds and one may check that this is the case by the exhaustive set of examples associated with CYBE discussed in \cite{Araujo:2017jap}. For the mCYBE case, we show this explicitly in the next section.

\subsection{$I$ and pure spinor formulation}

In the previous section we gave two meanings to the $I$ vector in the open string frame: we derived \eqref{div_theta} and argued that $I$ may be viewed as a current for the center of mass $U(1)$ gauge symmetry that we are left with after fixing the $\Lambda$-gauge. In \cite{Wulff:2016tju} it was suggested that the vector field $I$ may be related to some non-physical states that appear in the pure spinor analysis of supergravity \cite{Bedoya:2010qz, Mikhailov:2012id, Mikhailov:2014qka}. In this section we give additional evidence for this relation.

Type IIB string worldsheet theory can be characterized by the following data \cite{Berkovits:2001ue}: 
i) A conformally invariant action $S_{IIB}$, 
ii) a pair of nilpotent BRST operators $Q_L$ and $Q_R$ satisfying
	\be 
	Q^2_L = Q^2_R=\{ Q_L, Q_R\} = 0
	\ee
	and a grading, the \emph{ghost number}, for which $\mathtt{gh}(Q_L)= \mathtt{gh}(Q_R)=+1$,
iii) and a pair of $b$-ghosts $b$ and $\bar b$ satisfying $\{Q, b\} = T$ and $\{Q, \bar b\} = \bar T$,
	where $Q=Q_L + Q_R$. Let us now assume the existence of a theory satisfying these axioms.

Infinitesimal deformations parametrized by the \emph{integrated vertex operators} $V_i^{(2)}$  are defined by 
\be 
\label{deformation}
\begin{split} 
S & = S_{IIB} + \eta \int V_1^{(2)} + \eta^2 \int V_2^{(2)} + \cdots\\
Q & = Q + \eta Q_1 + \eta^2 Q_2 + \cdots
\end{split}
\ee
For sigma models on $\mathfrak{g}=\mathfrak{psu}(2,2|4)$, the $\beta$-deformations \cite{Lunin:2005jy} are described by vertex operators,
\be 
\label{vertex}
V_1^{(2)} = \frac{1}{2}r^{ab}j_a\wedge j_b\; ,
\ee
where $j_a$ are conserved currents of global symmetries, $r^{ab}$ are components of the tensor $r\in (\mathfrak{g}\wedge \mathfrak{g})_0/\mathfrak{g}$, and $(\mathfrak{g}\wedge \mathfrak{g})_0$ stands for the subspace of $(\mathfrak{g}\wedge \mathfrak{g})$ generated by objects of the form $a\wedge b$ with $[a,b]=0$. An important feature in the present analysis is that the second order expansion in $\eta$ implies that $r$ satisfies the homgeneous CYBE. The simplest example is given by the case when $r\in su(4)$.  As we know, this deformation gives the Lunin-Maldacena background \cite{Lunin:2005jy} that can be obtained by a TsT transformation. See \cite{Bedoya:2010qz} for further details.

It is easy to write a perturbative expansion similar to the deformations by integrated vertex (\ref{deformation}) for the YB deformed $\sigma$-model action. Using the projector $P_-^{\a\b}=\frac{1}{2}(\gamma^{\a\b}-\epsilon^{\a\b})$ where $\gamma_{\a\b}$ is the worldsheet metric, the YB deformations \cite{Delduc:2013qra, Kawaguchi:2014qwa} of AdS$_5\times S^5$ are defined by
\be 
\label{yb-def}
{\cal L} = \frac{(1 + c \eta^2)^2}{2(1-c\eta^2)}P^{\a \b}_- \textrm{Str}\left(A_\a \circ \dd \circ \frac{1}{1-\eta R\circ \dd}\circ A_\b \right)\; ,
\ee
where
\be 
\dd \equiv P_1 +\frac{2}{1-c\eta^2}P_2 - P_3\; ,
\ee
and $P_i$ are projectors onto the $\mathbb{Z}_4$ grading of the algebra $\mathfrak{psu}(2,2|4)$, with $c=0$ for deformations based on the homogeneous CYBE \cite{Kawaguchi:2014qwa} and 
$c=1$ for the mCYBE considered in \cite{Delduc:2013qra}.  

The perturbative expansion for the case $c=0$ is obvious: since the perturbation by integrated vertex in the pure spinor formalism (\ref{deformation}) is related to the homogeneous CYBE at ${\cal O}(\eta^2)$, it is clear that both pictures must be physically equivalent.

In what follows, the analysis of the $I$ field refers to deformations corresponding to homogeneous CYBE solutions, but we briefly discuss what we can expect from the analysis of deformations built from mCYBE solutions. In this case, given an infinitesimal parameter $\eta$ we have the expansion
\begin{widetext}
\be 
\begin{split}
{\cal L} & = -\frac{1}{2}P_-^{\a\b} \textrm{Str} \left(A_\a \circ \dd_0 \circ A_\b \right) - \frac{1}{2}\eta P_-^{\a\b} \textrm{Str} \left[A_\a \circ ( \dd_0 \circ R  \circ \dd_0) \circ A_\b \right]\\
& \frac{1}{2}\eta^2 P_-^{\a\b} \textrm{Str} \left[ A_\a \circ( \dd_0 \circ R \circ \dd_0 \circ R \circ \dd_0 + 2 P_2 + 3 \dd_0) \circ A_\b \right] + {\cal O}(\eta^3)
\end{split}
\ee
\end{widetext}
where $\dd_0 \equiv \dd\ |_{\eta=0}$ and the first term is the Metsaev-Tseytlin action \cite{Metsaev:1998it}.  Therefore, based on our previous experience, we may conjecture that there exists a perturbation by a vertex in the pure spinor formalism of the form 
\be 
\widetilde V_1^{(2)} = \frac{1}{2}\tilde{r}^{ab}j_a\wedge j_b\; , 
\ee 
where $\tilde{r}$ satisfies the mCYBE. All in all, we see that these two deformations pictures are consistent with each other, provided this mild conjecture is satisfied.

Now, let us return to the homogeneous CYBE picture. Consistency with supergravity imposes the additional condition on the vertex operator (\ref{vertex}): 
\be 
\label{unimodularity}
r^{ab}f_{ab}^{\phantom{ab} c} = 0\; ,
\ee
where $f_{ab}^{\phantom{ab} c} $ are the structure constants of the Lie algebra $\mathfrak{g}$. The authors of \cite{Bedoya:2010qz} have shown that when this condition is not satisfied, there are vertex operators associated with states, we call \emph{$\eta$-states} \footnote{\emph{Non-physical states} in their terminology.}, which are not present in the type IIB spectrum. 

In \cite{Mikhailov:2012id}, Mikhailov considers the flat space limit of the deformation (\ref{deformation}) and performs an analysis of the $\eta$-states. But since his analysis is in flat space and at first order in the $\eta$-expansion, there are many problems to be solved for finite transformations and curved backgrounds. There is one particular open question that can be solved using the connection between vertex operators and YB deformations: \emph{Do the $\eta$-states survive at all perturbative orders in $\eta$ in curved backgrounds?} 

From the YB deformation viewpoint, the condition (\ref{unimodularity}),  called \emph{unimodularity condition}, appears in \cite{Borsato:2016ose} and implies that deformed backgrounds satisfy the type II supergravity equations of motion, if and only if (\ref{unimodularity}) is satisfied. Failure of the unimodularity condition (\ref{unimodularity}) means that these YB deformed backgrounds satisfy the so called \emph{generalized supergravity equations of motion}, that take into account a covector $X$ with components
\be
\label{X} 
X_{M} \equiv \partial_{M} \Phi + ( g_{ N M} + B_{N M} ) I^{N},
\ee
where $I^M$ is a Killing vector in the generalized type II background. Furthermore, in the special case $I=0$, we get $X=\partial \Phi$, which can be interpreted as the gradient of the dilaton field $\Phi$. In this case, we recover the usual type II supergravity equations of motion.

Therefore, just one of the $\eta$-states in \cite{Bedoya:2010qz} survives at all perturbative orders in $\eta$, and it should correspond to the vector field $I$ in the YB deformation context. In order to make this correspondence more explicit, we follow a simple strategy: we know that type IIB supergravity does not have any vector field in the physical spectrum, then we should keep track of states that give origin to vectors in the BRST cohomology of \cite{Mikhailov:2014qka}. Just one of these vector-like states may be realized as a  descendant of a physical type IIB state, namely, the gradient of the dilaton $V= \partial \Phi$. 

As usual, kappa-symmetry plays the crucial role restricting the extra degrees of freedom in a supergravity theory, including the generalized supergravity in the absence of Weyl invariance of the worldsheet  theory. In pure spinor formalism, the correct number of degrees of freedom is carried and imposed through BRST charges and one would hence expect to find a counterpart of $I$ through BRST cohomology analysis.

The BRST cohomology in the flat space limit \footnote{This corresponds to a Taylor expansion of the AdS${}_5\times $S${}_5$ solution around a point $x_\ast$ \cite{Mikhailov:2012id}} and at order $\eta$ has been performed in \cite{Mikhailov:2014qka}. In our analysis, the essential states in this cohomology are the covectors $A^+$ and $A^-$, which satisfy the conditions
\be 
\label{A-eq}
\begin{split}
\partial_{[M}A_{N]}^+ & = 0\\ 
\partial_{(M}A_{N)}^- & = 0\; .
\end{split}
\ee
The vector field $A^+$ has been identified as the gradient of the dilaton, that is
\be 
A_M^+:=\partial_M \Phi\; .
\ee

The state $A^-$ is clearly an $\eta$-state, as there is no additional vector field in the type IIB string spectrum. Furthermore, through some mild assumptions, it is reasonable to conjecture that this state is linearly related to the vector $I$, that is $I \sim A^-$.

In the undeformed limit $\eta=0$, we recover a type IIB supergravity solution, where $I=0$. Therefore, the perturbative expansion of this field must be of the form $I=\eta I^{(1)} + \eta^2 I^{(2)} + {\cal O}(\eta^3)$. In addition, we already know that the vector field $I$ is a Killing vector in the generalized supergravity solution, and that the flat space limit of the Killing vector equation is simply
\be 
\label{killveceq}
\nabla_M I_N + \nabla_N I_M = 0\ \ \xrightarrow[\texttt{flat space}] \ \ \partial_M I_N + \partial_N I_M = 0\; ,
\ee
which is, obviously, the second equation in (\ref{A-eq}). Therefore, it is easy to see that the $\eta$-state $A^-$ corresponds to the first order component of $I$ in its perturbative expansion, that is
\be 
\label{Ivect}
A_M^- = I_M^{(1)}\; .
\ee
Using this perturbative analysis, we can conjecture that in a complete BRST analysis all states of $\eta$-states combine to give the Killing vector $I$ in the Yang-Baxter deformation context. Additionally, all other $\eta$-states in the BRST cohomology are obstructed by higher orders corrections and are, using the terminology of \cite{Bedoya:2010qz}, truly \emph{non-physical states}. Finally, equations (\ref{killveceq}) and (\ref{Ivect}) realize explicitly the observation of \cite{Mikhailov:2014qka} that $\eta$-states are in correspondence with global symmetries of the background. Evidently, the unarguable proof for this identification requires methods to calculate the complete BRST cohomology for finite deformations in the pure spinor formalism.

\section{$\eta$-deformations and mCYBE } 
In previous work \cite{Araujo:2017jkb, Araujo:2017jap}, we have largely discussed YB deformations based on $r$-matrix solutions to the homogeneous CYBE. For the analysis of the bulk D3-brane action, we were ambivalent to whether the $r$-matrix solved the homogeneous CYBE or mCYBE. In principle, we could check $Q_5$ for all the known $\eta$-deformations \cite{Delduc:2014kha}, but apart from the ABF solution \cite{Arutyunov:2013ega, Arutyunov:2015qva}, where $Q_5$ coincides with the undeformed five-form flux, the explicit form of the RR fluxes are not known. So, here we will adopt a different tack and demonstrate that equation (\ref{div_theta}) is valid for all known $\eta$-deformations. We will see that pure gauge contributions to the NSNS two-form play a special role.

To be clear, our approach is to take $g$ and $B$, which can be read off from the string $\sigma$-model \cite{Arutyunov:2013ega, Delduc:2014kha}, determine $\Theta$ and then use equation (\ref{div_theta}) to work out $I$, which we will directly compare with the Killing vector from the generalized supergravity solution. For the YB deformations identified by Delduc et al. \cite{Delduc:2014kha}, we will determine both $I$ and $\Phi$ and check that the dilaton EOM \cite{Arutyunov:2015mqj}, 
\be
\label{dilaton_EOM}
R - \frac{1}{12} H_{M N P} H^{M N P} + 4 \nabla_{M} X^{M} - 4 X_{M} X^{M} = 0, 
\ee
with $X$ defined in (\ref{X}), is satisfied. This provides a powerful consistency check on our results.  

We recall the $\eta$-deformed geometry for AdS$_5\times$S$^5$ \cite{Arutyunov:2013ega}, 
\begin{widetext}
\bea
\label{ABF}
\dd s^2  &=& - \frac{(1+\rho^2) \dd t^2}{1-\kappa^2 \rho^2}  + \frac{\dd \rho^2}{( 1+ \rho^2) (1-\kappa^2 \rho^2)} + \frac{\rho^2 \dd \zeta^2}{1+ \kappa^2 \rho^4 \sin^2 \zeta}
+ \frac{\rho^2 \cos^2 \zeta \dd \psi_1^2}{1+ \kappa^2 \rho^4 \sin^2 \zeta} + \rho^2 \sin^2 \zeta \dd \psi_2^2 \nn
&+& \frac{(1-r^2) \dd \phi^2}{1 + \kappa^2 r^2} + \frac{\dd r^2}{(1-r^2)(1+\kappa^2 r^2)} + \frac{r^2 \dd \xi^2}{1+ \kappa^2 r^4 \sin^2 \xi} + \frac{ r^2 \cos^2 \xi \dd \phi_1^2}{1 + \kappa^2 r^4 \sin^2 \xi} + r^2 \sin^2 \xi \dd \phi_2^2,  \\
B &=& - \frac{\kappa  \rho^4 \sin 2 \zeta}{2(1 + \kappa^2 \rho^4 \sin^2 \zeta)} \dd \zeta \wedge \dd \psi_1 + \frac{\kappa \rho}{1- \kappa^2 \rho^2} \dd t \wedge \dd \rho
-  \frac{\kappa r^4 \sin 2 \xi}{2 (1 + \kappa^2 r^4 \sin^2 \xi)} \dd \phi_1 \wedge \dd \xi +  \frac{\kappa r }{1 + \kappa^2 r^2} \dd \phi \wedge \dd r. \nonumber
\eea
\end{widetext}
Relative to its original incarnation, we have retained the pure gauge terms in the $B$-field (see also \cite{Delduc:2014kha}). It turns out that these terms are important for the open string metric to be undeformed. It is worth noting that there is a simple analytic continuation that takes the AdS$_5$ deformation to the S$^5$ deformation. Interestingly, just as for $\lambda$-deformations \cite{Sfetsos:2014cea, Demulder:2015lva}, it can be shown that the dilaton equation forces the same deformation parameter in AdS$_5$ and S$^5$. Since $\eta$-deformations correspond to scaling limits of $\lambda$-deformations \cite{Hoare:2015gda}, this is not too surprising.   

From (\ref{Theta-def}), it is an easy exercise to determine $\Theta$:
\bea
\label{theta_ABF}
\Theta^{\zeta \psi_1 } &=& \kappa \tan \zeta, \quad \Theta^{t \rho} = \kappa \rho, \nn
\Theta^{\xi \phi_1} &=& -\kappa \tan \xi, \quad \Theta^{\phi r } = - \kappa r,  
\eea
and $I$ using (\ref{div_theta}), 
\be
\label{vec_ABF}
I = \kappa ( - 4 \partial_t + 2 \partial_{\psi_1} + 4 \partial_{\phi} - 2 \partial_{\phi_1}), 
\ee
which precisely agrees with the ABF result \cite{Arutyunov:2015qva}. In contrast to ABF, the pure gauge terms in the $B$-field mean the dilaton is different, though $X$ of course is the same. 

To determine the dilaton, we will now introduce a simple trick that has been largely overlooked. It is well-known that $e^{-2 \Phi} \sqrt{g}$ is an invariant of the Buscher rules of T-duality. Existing results extend this to homogeneous YB deformations \cite{Araujo:2017jap}, and having verified that the open string metric is undeformed for the ABF solution, it follows that this is also an invariant of $\eta$-deformations.
This allows us to easily determine the dilaton for the ABF solution (\ref{ABF}), 
\be
e^{-2 \Phi} =  (1- \kappa^2 \rho^2) (1 + \kappa^2 \rho^4 \sin^2 \zeta) (1+ \kappa^2 r^2) (1+ \kappa^2 r^4 \sin^2 \xi). \nonumber
\ee
For consistency, we have checked the dilaton equation (\ref{dilaton_EOM}). 

Emboldened by this success, we proceed to analyze the $\eta$-deformations identified in \cite{Delduc:2014kha}. Note, for these geometries the corresponding $I$, $\Phi$ and RR sector are not known. We proceed as before: we check that open string metric is undeformed in the presence of pure gauge contributions \footnote{This allows us to identify and correct typos in \cite{Delduc:2014kha}.}, determine the dilaton from the invariant and deduce the Killing vector from (\ref{div_theta}). In each case, we will show that the picture is complete by checking the dilaton EOM. One special feature of the $\eta$-deformations \cite{Delduc:2014kha} is that the $r$-matrix is a permutation of the ABF $r$-matrix and only the AdS$_5$ part of the deformation differs from the ABF solution. 

Focusing on the AdS$_5$ deformation, which is new relative to ABF, the first $\eta$-deformation may be written as \cite{Delduc:2014kha} 
\begin{widetext}
\bea
\dd s^2  &=& - \frac{(1+\rho^2)}{s} \dd t^2 + \frac{1 + \kappa^2 \sin^2 \zeta - \kappa^2 \rho^2 ( 1+ \rho^2) \cos^2 \zeta \sin^2 \zeta}{(1+\rho^2) f s } \dd \rho^2 
+ \frac{\rho^2 ( 1+ \kappa^2 (1+\rho^2) \cos^2 \zeta - \kappa^2 \rho^2 \sin^4 \zeta)}{f s } \dd \zeta^2 \nn &+& \frac{2 \kappa^2 \rho( 1+ \rho^2 \sin^2 \zeta) \cos \zeta \sin \zeta}{f s} \dd \rho \dd \zeta 
+ \frac{\rho^2 \cos^2 \zeta}{f} \dd \psi_1^2 + \rho^2 \sin^2 \zeta \dd \psi_2^2, \nn
B &=&  - \frac{1}{2 \kappa} \dd \log f \wedge \dd \psi_1 -  \frac{\kappa}{s} \left( \rho \sin^2 \zeta \dd \rho + \rho^2 (1+ \rho^2) \sin \zeta \cos \zeta \dd \zeta \right) \wedge \dd t. 
\eea
\end{widetext}
where we have introduced the functions:
\bea
f(\rho, \zeta) &=& 1 + \kappa^2 + \kappa^2 \rho^2 \cos^2 \zeta, \nn s(\rho, \zeta) &=& 1 - \kappa^2 \rho^2 (1+\rho^2 \cos^2 \zeta) \sin^2 \zeta, \nn
h(\rho, \zeta) &=& 1 + \kappa^2 ( 1+ \rho^2) + \kappa^2 \rho^2 (1+ \rho^2) \cos^2 \zeta. 
\eea
The open string metric is undeformed, as expected, and $\Theta$ takes the form: 
\bea
\Theta^{t \rho } &=& \kappa \rho \sin^2 \zeta, \quad \Theta^{t \zeta} = \kappa \cos \zeta \sin \zeta, \nn
\Theta^{\rho \psi_1} &=& \kappa (\rho^{-1} + \rho), \quad \Theta^{\zeta \psi_1} = - \kappa \rho^{-2} \tan \zeta. 
\eea
Including the contribution from the internal space, the Killing vector may be determined from (\ref{div_theta}),
\be
I =  \kappa ( - 2 \partial_t + 4 \partial_{\psi_1}  + 4  \partial_{\phi} - 2  \partial_{\phi_1} ). 
\ee
Finally, the dilaton can be calculated from the invariant $e^{-2 \Phi} \sqrt{ g}$: 
\be
e^{-2 \Phi} = f  s (1+ \kappa^2 r^2) (1+ \kappa^2 r^4 \sin^2 \xi). 
\ee
To confirm that our treatment is correct, we have checked that the dilaton EOM (\ref{dilaton_EOM}) is satisfied. 

Finally, we consider the remaining $\eta$-deformation \cite{Delduc:2014kha}: 
\begin{widetext}
\bea
\dd s^2 &=& - (1+\rho^2) \dd t^2  + \frac{1+ \kappa^2 + \kappa^2 \rho^2 (2 + \rho^2) \cos^2 \zeta}{(1+ \rho^2) f h} \dd \rho^2  + \frac{\rho^2 ( 1+ \kappa^2 (1+ \rho^2))}{f h } \dd \zeta^2 - \frac{\kappa^2 \rho^3 \sin(2 \zeta)}{f h} \dd \rho \dd \zeta \nn
&+& \frac{\rho^2 \cos^2 \zeta}{f} \dd \psi_1^2 + \frac{\rho^2 \sin^2 \zeta}{h} \dd \psi_2^2, \nn
B &=& - \frac{1}{2 \kappa} \dd \log f \wedge \dd \psi_1  - \frac{\kappa}{ h} \left( \rho \sin^2 \zeta \dd \rho + \rho^2 (1+ \rho^2) \sin \zeta \cos \zeta \dd \zeta \right) \wedge \dd \psi_2.
\eea
\end{widetext}
One can check that the open string metric is again undeformed and the external bivector takes the form, 
\bea
\Theta^{\rho \psi_i} &=&  \kappa ( \rho^{-1} + \rho), \quad 
\Theta^{\zeta \psi_1 } = - \kappa \rho^{-2} \tan \zeta, \nn \Theta^{\zeta \psi_2} &=& \kappa \rho^{-2} ( 1+ \rho^2) \cot \zeta. 
\eea
Including the internal directions, the Killing vector is 
\be
I =  \kappa ( 4  \partial_{\psi_1} + 2 \partial_{\psi_2} +4 \partial_{\phi} - 2 \partial_{\phi_1}), 
\ee
and the dilaton is 
\be
e^{-2 \Phi} = f  h (1+ \kappa^2 r^2) (1+ \kappa^2 r^4 \sin^2 \xi). 
\ee
Again, we have checked the dilaton equation.

We end this section with some comments. It is easy to check that the Jacobi identity for the bivector, or more concretely,  
\be
\label{Jacobi}
\Theta^{[L P} \partial_{P} \Theta^{M N]} = 0, 
\ee
is satisfied for the three examples of $\eta$-deformations we have considered. We observe that this is a necessary condition for the bivector $\Theta$ to correspond to an NC parameter. 

Furthermore, in \cite{Araujo:2017jap} it was noted that the bivector $\Theta$ exhibits holographic noncommutativity: although it may depend on the holographic direction $z$, working in Poincar\'e patch with the boundary at $z=0$, the dependence on the holographic direction drops out consistently at the boundary, i. e. at the boundary both legs of $\Theta$ are along the boundary, $\Theta^{\mu z}|_{z=0} =0$. To see that the same picture applies equally well to the ABF solution (\ref{ABF}), one can employ the following map from (unit radius) global AdS$_5$ to the Poincar\'e metric, 
\bea
\rho &=& \frac{1}{2 z} \sqrt{ 4 \vec{x}^2 + (z^2-1 + \vec{x}^2 -t^2)^2}, \nn
\cos \tau &=& \frac{z^2+1 + \vec{x}^2 -t^2}{\sqrt{4 t^2 + (z^2+1 + \vec{x}^2 -t^2)^2}}, \nn
\cos \zeta &=& \frac{2 \sqrt{ (x_1^2 + x_2^2)}}{\sqrt{ 4(x_1^2 + x_2^2  + x_3^2) + (z^2-1 + \vec{x}^2 -t^2)^2}}, \nn
\cos \psi_1 &=& \frac{x_1}{\sqrt{x_1^2 + x_2^2}}, \nn
\cos \psi_2 &=& \frac{2 x_3}{\sqrt{4 x_3^2 + (z^2-1 + \vec{x}^2 -t^2)^2}}, 
\eea
where $\vec{x} \equiv (x_1, x_2, x_3)$. Evaluating the ABF bivector (\ref{theta_ABF}) in Poincar\'e coordinates at the boundary, $\Theta^{\mu \nu} = \Theta^{MN} |_{z=0}$, we have confirmed that all $z$-dependence disappears leaving the following non-zero contributions to the bivector in Poincar\'e patch: 
\bea
\Theta^{t x_1} &=&  - \kappa \frac{t x_2}{2} (x_{\mu} x^{\mu}-1), \quad
\Theta^{t x_2} =  \kappa \frac{t x_1}{2} (x_{\mu} x^{\mu}-1), \nn
\Theta^{x_1 x_2} &=& -\kappa \left( x_3^2 + \frac{1}{4} (x_{\mu} x^{\mu}-1) (x_{\mu} x^{\mu} - x_1^2 -x_2^2 -1) \right), \nn 
\Theta^{x_1 x_3} &=& \kappa \frac{x_2 x_3}{2} (x_{\mu} x^{\mu}+1), \nn
\Theta^{x_2 x_3} &=& -\kappa \frac{x_1 x_3}{2} (x_{\mu} x^{\mu}+1).
\eea
We expect a similar outcome for other $\eta$-deformations. The corresponding expression for the Killing vector $I$ can be written in terms of the Killing vectors for the Poincar\'e metric of AdS$_5$ as 
\be
I = 2 \kappa ( P_0 + K_0 +M_{12}), 
\ee
where $P_0$ and $K_0$ denote translations and special conformal transformations in the temporal direction, respectively, and $M_{12}$ is a rotation in the $(x_1, x_2)$-plane. We refer the reader to (A.2) of \cite{Araujo:2017jap} for details of the notation. Note, the holographic behavior of $I$ is dictated by (\ref{div_theta}).

\section{Conclusions}
One of the results of this letter is completing the derivation of equation (\ref{div_theta}) from the requirement that the action of D3-branes is invariant under $\Lambda$-symmetry. An interesting feature of the calculation is that the equations of generalized supergravity are incorporated through a bulk AdS$_5$ Wess-Zumino term, which couples the D3-brane to the background. This provides an explanation for the relation, which was observed to hold for all homogeneous Yang-Baxter deformations on a case by case basis. In this letter we show that $\eta$-deformations have an open string metric that is undeformed and satisfy the same equation. This allows us to interpret $I$ as a current for the center of mass motion on the worldvolume of the D3-brane. 

As a secondary result, we noted that the Bianchi identities and EOMs of generalized supergravity take a strikingly simple form when expressed in terms of Page forms. We conjectured that the non-zero Page forms, as well as associated Page charges  are invariants of YB deformations. We showed that there always exist suitably chosen pure gauge contributions to the NSNS $B$-field, such that open string metric and consequently $e^{-2 \Phi} \sqrt{g}$ remains undeformed for all (known) YB deformations. This may be used as a handy way to determine the dilaton for all YB deformations, including $\eta$-deformations.

Lastly, using the deformation of the pure spinor action through integrated vertex operators, we have identified the states in the BRST cohomology that give origin to the vector field $I$ in the $\eta$-deformation context. As a by-product, we have shown that other non-physical states in the BRST cohomology of \cite{Mikhailov:2014qka} are suppressed by higher order corrections in the curvature and/or deformation parameter $\eta$.

It is a straightforward exercise to extend the analysis presented here to $\eta$-deformations of AdS$_3 \times$S$^3$ and AdS$_2 \times$S$^2$ geometries \cite{Hoare:2014pna, Lunin:2014tsa}. Once pure gauge contributions to the NSNS two-form are introduced, we expect the open string metric to be undeformed and for equation (\ref{div_theta}) to hold with the slight story twist that it will arise from variations of the actions of D1-branes, D5-branes, intersecting D3-branes, etc.

One problem that has yet to be unraveled is the interpretation of the $\eta$-deformed geometries \cite{Delduc:2013qra, Arutyunov:2013ega} in the context of AdS/CFT. This problem is thorny, since in contrast to simpler homogeneous YB deformations, one has to consider both AdS$_5$ and S$^5$ deformations. As a result, there is less residual global symmetry,  but the consensus is that the dual gauge theory, if it exists, is expected to be some NC deformation of $\mathcal{N}=4$ super Yang-Mills. In this letter, we have extracted a candidate NC parameter for the ABF solution \cite{Arutyunov:2013ega}, shown that it is associative in the sense that the Jacobi (\ref{Jacobi}) is satisfied, which means that it is consistent with a non-commuting open string description. Moreover, we have checked in Poincar\'e patch that the holographic direction decouples at the boundary, where the putative dual gauge theory is expected to reside.

\section*{Acknowledgements}
We thank I. Bakhmatov, D. Berman, M. Duff, B. Hoare and D.C. Thompson for discussion on related projects. E. \'O C acknowledges the kind hospitality of the workshop ``Recent Advances in T/U-dualities and Generalized Geometries", June 6-9 2017, Zagreb, where this work was initiated. 
The work of J.S.\ is supported by the Japan Society for the Promotion of Science (JSPS). 
 M.M. Sh-J. is supported in part by the Saramadan Iran  Federation, the junior research chair on black hole physics by the Iranian NSF 
and the ICTP Simons Associate fellowship and ICTP network project NT-04 K.Y.\ acknowledges the Supporting Program for Interaction-based Initiative Team Studies (SPIRITS) 
from Kyoto University and a JSPS Grant-in-Aid for Scientific Research (C) No.\,15K05051. 
This work is also supported in part by the JSPS Japan-Russia Research Cooperative Program.  

\appendix
\section{Review of Generalized Supergravity}
The equations of motion for the metric, NSNS two-form and dilaton in the generalized type IIB supergravity are given by 
\bea
 &&\scalebox{0.9}{$\displaystyle
R_{MN} - \frac{1}{4}\,H_{MPQ}\,H_N{}^{PQ} + D_M X_N + D_N X_M =T_{MN} \,,$} 
\nn
 &&\scalebox{0.9}{$\displaystyle
\frac{1}{2}\,D^K H_{KMN} +\frac{1}{2}\mathcal{F}^K\mathcal{F}_{KMN}+\frac{1}{12}\mathcal{F}_{MNKLP}\mathcal{F}^{KLP}$}\nn
&&\scalebox{0.9}{$\displaystyle\qquad= X^K H_{KMN} + D_M X_{N} - D_N X_M \,,$}
\nn
 &&\scalebox{0.9}{$\displaystyle
R - \frac{1}{12}\, H_{MNP}H^{MNP} + 4\,D_M X^M - 4\,X^M X_M = 0 \,,$}
\label{GSE-NSNS}
\eea
where the spacetime indices $M,N,\ldots$ run from 0 to 9, $H$ is the field strength of the NSNS two-form. The appearance of a vector field $X$ indicates the generalization. Note here that the dilaton $\Phi$ is implicitly included in $X$ with a spacetime derivative and an extra vector field $I$ as explained later.  
The flux contribution $T_{MN}$ is defined as
\bea
\scalebox{0.8}{$\displaystyle
T_{MN}$} &\equiv& 
\scalebox{0.8}{$\displaystyle\frac{1}{2}\mathcal{F}_M\mathcal{F}_N
+\frac{1}{4}\mathcal{F}_{MKL}\mathcal{F}_N{}^{KL}
+\frac{1}{4\times 4!}\mathcal{F}_{MPQRS}\mathcal{F}_N{}^{PQRS}$} \nn
&&\scalebox{0.8}{$\displaystyle
- \frac{1}{4}G_{MN} (\mathcal{F}_K\mathcal{F}^{K}
+\frac{1}{6}\mathcal{F}_{PQR}\mathcal{F}^{PQR})\,.$}
\eea
Here $\mathcal{F}_M\,,\mathcal{F}_{MNK}\,,\mathcal{F}_{MNKPQ}$ are 
the rescaled RR field strengths, i. e.  
\be
\mathcal{F}_{n_1n_2\ldots}={\rm e}^{\Phi}F_{n_1n_2\ldots}\,. 
\ee
The vector $X$ is decomposed as
\be
X_M\equiv I_M +Z_M\,,
\ee
where $I$ and $Z$ satisfy relations
\bea
 &&D_M I_N + D_N I_M =0\,, \\
 &&I^K\, H_{KMN} + D_M Z_N - D_N Z_M =0\,,\\
 &&I^M\,Z_M = 0\,, 
\label{GSE-constraints}
\eea 
where $D$ is the closed string covariant derivative. 
Assuming that the Lie derivative of the NSNS two-form along the $I$-direction vanishes, one obtains the following expression:
\begin{eqnarray}\label{Z-solution}
Z_M = \partial_M \Phi - B_{MN} I^N\,.
\end{eqnarray}
We note that while the $X$ and hence the $Z$ fields are invariant nder $\Lambda$-gauge transformations, the specific solution \eqref{Z-solution} is not. This is written in a specific gauge. This gauge is the same as the one we used in our analysis of $\Lambda$-invariance of the DBI action.
Thus, in total, $X$ can be regarded as a composite field comprising the dilaton derivative, the NSNS two-form and 
the extra vector field $I$.

\end{document}